\documentclass[usenatbib,usegraphicx,useAMS,usenatbib]{pasa}
\usepackage{color}
\usepackage{graphicx}
\usepackage[caption=false]{subfig}
\usepackage{amsmath}

\title[An Isolated Star?]{A search for High Mass Stars Forming in Isolation using CORNISH \& ATLASGAL}
\author[Tremblay et al.]{Chenoa D. Tremblay$^{1}$\thanks{E-mail:
chenoa.tremblay@postgrad.curtin.edu.au}, Andrew J. Walsh$^{1}$, 
Steven N. Longmore$^{2}$, James S. Urquhart$^{3,4}$, and Carsten K\"onig$^{3}$
\\
\small $^{1}$International Centre for Radio Astronomy Research, Curtin University, GPO Box U1987, Perth WA 6845, Australia\\
\small $^{2}$Astrophysics Research Institute, Liverpool John Moores University, Twelve Quays House, Egerton Wharf, Birkenhead CH41 1LD, UK\\
\small $^{3}$Max-Planck-Institut f\"{u}r Radioastronomie, Auf dem H\"{u}gel 69, Bonn, Germany\\
\small $^{4}$Centre for Astrophysics and Planetary Science, University of Kent, Canterbury, CT2 7NH}

\jid{PASA}
\doi{10.1017/pas.\the\year.xxx}
\jyear{\the\year}

\begin{document}




\begin{abstract}
Theoretical models of high mass star formation lie between two extreme scenarios. At one extreme, all the mass comes from an initially gravitationally-bound core. At the other extreme, the majority of the mass comes from cluster scale gas, which lies far outside the initial core boundary. One way to unambiguously show high mass stars can assemble their gas through the former route would be to find a high mass star forming in isolation. Making use of recently available CORNISH and ATLASGAL Galactic plane survey data, we develop sample selection criteria to try and find such an object. From an initial list of approximately 200 sources, we identify the high mass star forming region G13.384+0.064 as the most promising candidate. The region contains a strong radio continuum source, that is powered by an early B-type star. The bolometric luminosity, derived from infrared measurements, is consistent with this. However, sub-millimetre continuum emission, measured in ATLASGAL, as well as dense gas tracers, such as HCO$^{+}$(3-2) and N$_{2}$H$^{+}$(3-2) indicate that there is less than $\sim100$\,M$_\odot$ of material surrounding this star. We conclude that this region is indeed a promising candidate for a high mass star forming in isolation, but that deeper near-IR observations are required to put a stronger constraint on the upper mass limit of young, lower mass stars in the region. Finally, we discuss the challenges facing future studies in proving a given high mass star is forming in isolation.

\end{abstract}


\maketitle%

\section{Introduction}
High mass stars --  O or early B type star of sufficient mass to produce a Type II supernova \cite{Zinnecker} or  $>$8 M$_{\odot}$ \cite{Miettinen} -- dominate the energy cycles and chemical enrichment of galaxies. However, understanding the formation of high mass stars remains a challenge, and several different theoretical formation scenarios have been proposed \cite{Zinnecker, Tan14}. The general observational phases (i.e. formation of cold dense cloud, gravitational collapse of a hot core, accretion, and formation of ultracompact H{\scriptsize II}~ regions) are typically agreed upon, but the dominant physical processes and their relevant time scales are still under debate. The problems associated with our understanding are that it is difficult to observe the early stages of formation due to high dust extinction, the theoretical problem is complex and high mass stars are seldom, if ever, formed in isolation \cite{Zinnecker}.

Over the last few decades, high mass star formation theories have been discussed in the context of two extreme scenarios: that of the turbulent core  \cite{McKee03} and the competitive accretion \cite{Bonnell01, Bonnell06} scenarios. In the former, all the mass comes from an initially gravitationally-bound core. In the latter, the majority of the mass comes from cluster scale gas that is far outside the initial core boundary. Early debate suggested that stellar collisions \cite{Bonnell98} could be a potential creation mechanism, but this has largely been discounted due to the extremely high stellar densities required\footnote{Although see Izumi et al. \shortcite{Izumi} who suggest a star system in the extreme outer galaxy ($>$18\,kpc) may have been formed by large scale collision.}. More recent theoretical and simulation work, adding more physics (e.g. radiation pressure, ionisation) and overcoming previous limitations in numerical methods, have found that increasing the feedback, initial density fluctuations and turbulence leads to an increase in the fraction of the final stellar mass which comes directly from an initial gravitationally-bound core \cite{Krumholz07, Krumholz12, Hennebelle11, Commercon11, Myers13, Bonnell04, Peters10a}. The apparent dichotomy between the competitive and core accretion models seems to be less extreme, or at least the reasons for the dichotomy are now better understood.

Nevertheless, understanding whether the two extreme formation scenarios are viable routes for high mass stars to assemble their mass has important consequences for the host galaxy. If high mass stars can only form in the presence of an attendant cluster, the stellar initial mass function (IMF) will be sampled very differently than if high mass stars can (albeit rarely) form in isolation \cite{Bastian}. When averaged on galactic scales, this can make a dramatic difference in the number of (very) high mass stars, and hence the level of feedback, chemical enrichment etc.

Given the wider importance of how the IMF is sampled, many observational studies have tried to find evidence of high mass stars forming in isolation \cite{Bressert, Tout, Weidner09, Weidner05, Parker07, deWit, Schilbach}.   These previous searches have primarily focused on optical and infrared data to find young high mass stars with no lower mass young stars around them. However, as these high mass stars must already have cleared their natal gas cloud in order to be optically visible, it is very difficult to determine if they \emph{formed} at their present location, as opposed to having been ejected from their parent stellar nursery of lower mass stars.

In this paper we try a different approach, aiming to find very young high mass stars while they are still embedded in their natal gas cloud. While extinction makes it impossible to find these objects in the optical and near-IR, their prodigious luminosity and Lyman continuum flux means they should be conspicuous at far-IR wavelengths  and have bright cm-continuum, free-free emission. Even in the most optimistic scenarios, high mass stars forming in isolation are expected to be very rare, requiring large-area surveys to identify candidates. Thanks to an enormous effort from the Galactic observational community, Galactic plane surveys now exist across much of the electromagnetic spectrum at sufficient sensitivity and resolution to identify the majority of young high-mass star formation regions in the Galaxy. With the data now in hand, we aim to use a simple selection criterion to pick out the best candidates for young, high mass stars forming in isolation.

\section{Sample Selection}

For this first attempt to try and find examples of high mass stars forming in isolation, we used a series of data summarised by Urquhart et al. \shortcite{Urquhart} for ATLASGAL \cite{Schuller09} and CORNISH \cite{Purcell13}. Urquhart et al. \shortcite{Urquhart} noted that targeted surveys of compact and ultra-compact (UC) H{\scriptsize II} regions identified by infrared colours can be contaminated with intermediate mass young stellar objects (YSOs) and planetary nebulae (PNe). However, by incorporating radio astronomy data mixed into the identification process, this is no longer a concern. Intermediate mass stars do not show radio continuum emission, so would be eliminated in the cross match.  Planetary nebulae are not usually associated with dust that is bright enough to be detected by ATLASGAL. \footnote{There were some instances noted where PNe were identified in ATLASGAL by their mid-infrared emission.  However, they were removed from the source list and monte-carlo simulations agreed with their identified numbers.}

\begin{figure}
        \includegraphics[width=0.48\textwidth]{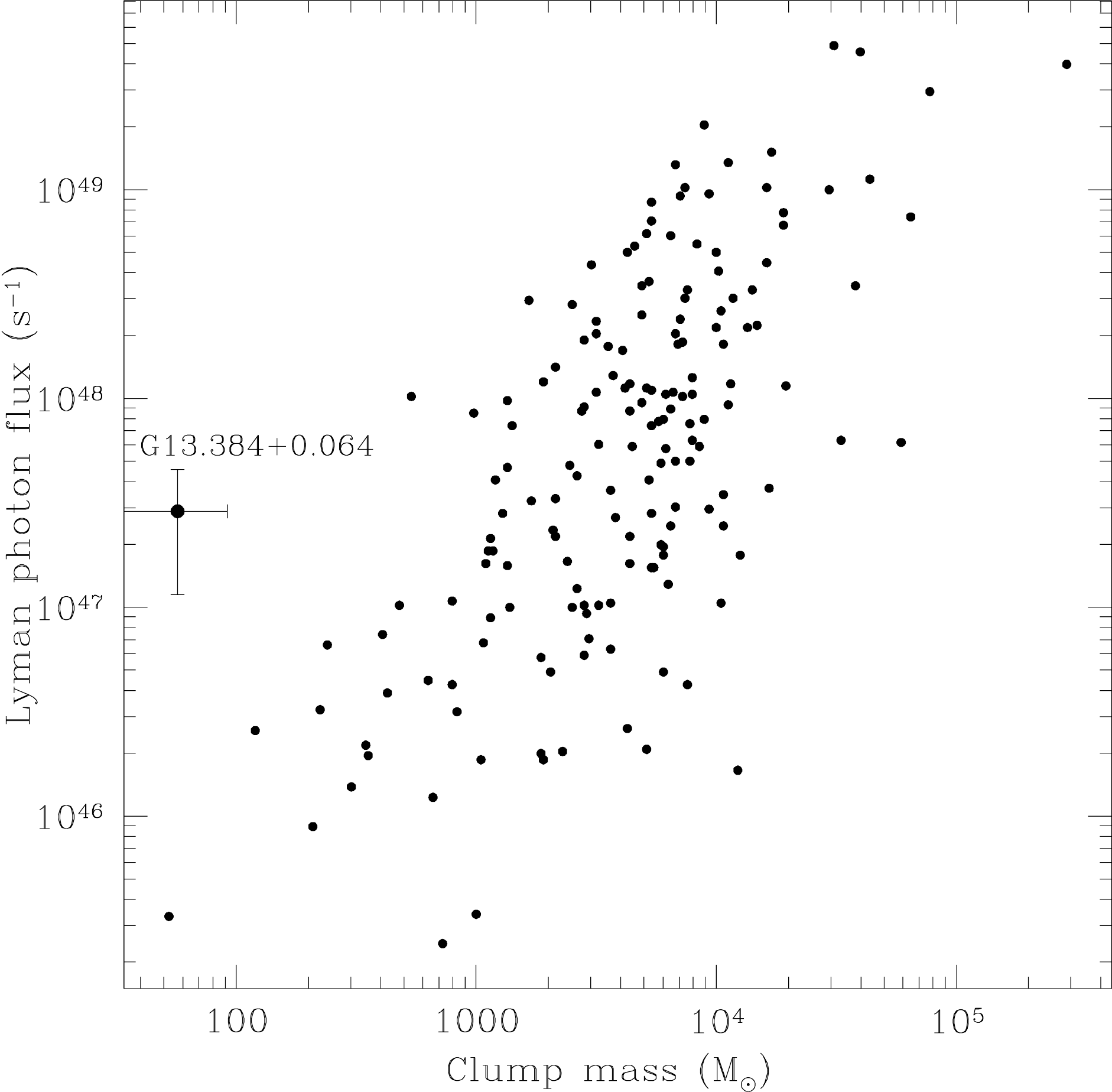}
        \caption{Plot of clump mass versus Lyman photon flux for a series of ATLASGAL and CORNISH sources, similar to the upper panel of Figure 26 in Urquhart et al.  \protect\shortcite{Urquhart}.  When the clump mass is plotted against the bolometric luminosity, the source at G13.384+0.064 does not stand out as different compared to other star forming regions.  This discrepancy is investigated in this paper.  The error bars represent the best available data reported in this paper. }
        \label{Andrewplot}
\end{figure}

We use their selection criteria to select approximately 200 bona fide YSOs, associated with compact and UC H{\scriptsize II} regions. Figure \ref{Andrewplot} compares the clump gas mass and Lyman continuum flux of these regions.  The source G13.384+0.064 stands out in Figure \ref{Andrewplot} as having a very low gas mass for its Lyman continuum flux. This Figure is closely matched to the upper panel from Figure 26 in Urquhart et al.  \protect\shortcite{Urquhart}. The aim of this paper is to better constrain the luminosity, mass and lyman photon flux based on a thorough literature search of the region and using GLIMPSE  \cite{Churchwell, Benjamin} and MIPSGAL \cite{Carey09} data to to test for evidence of a surrounding cluster. We selected this as the most promising candidate in our initial sample for a high mass star forming in isolation. Below we describe our efforts to use data in the literature to determine whether or not we can confirm or rule out this status.

\begin{figure*}
	\centering
	 {%
	 \includegraphics[width=0.48\textwidth]{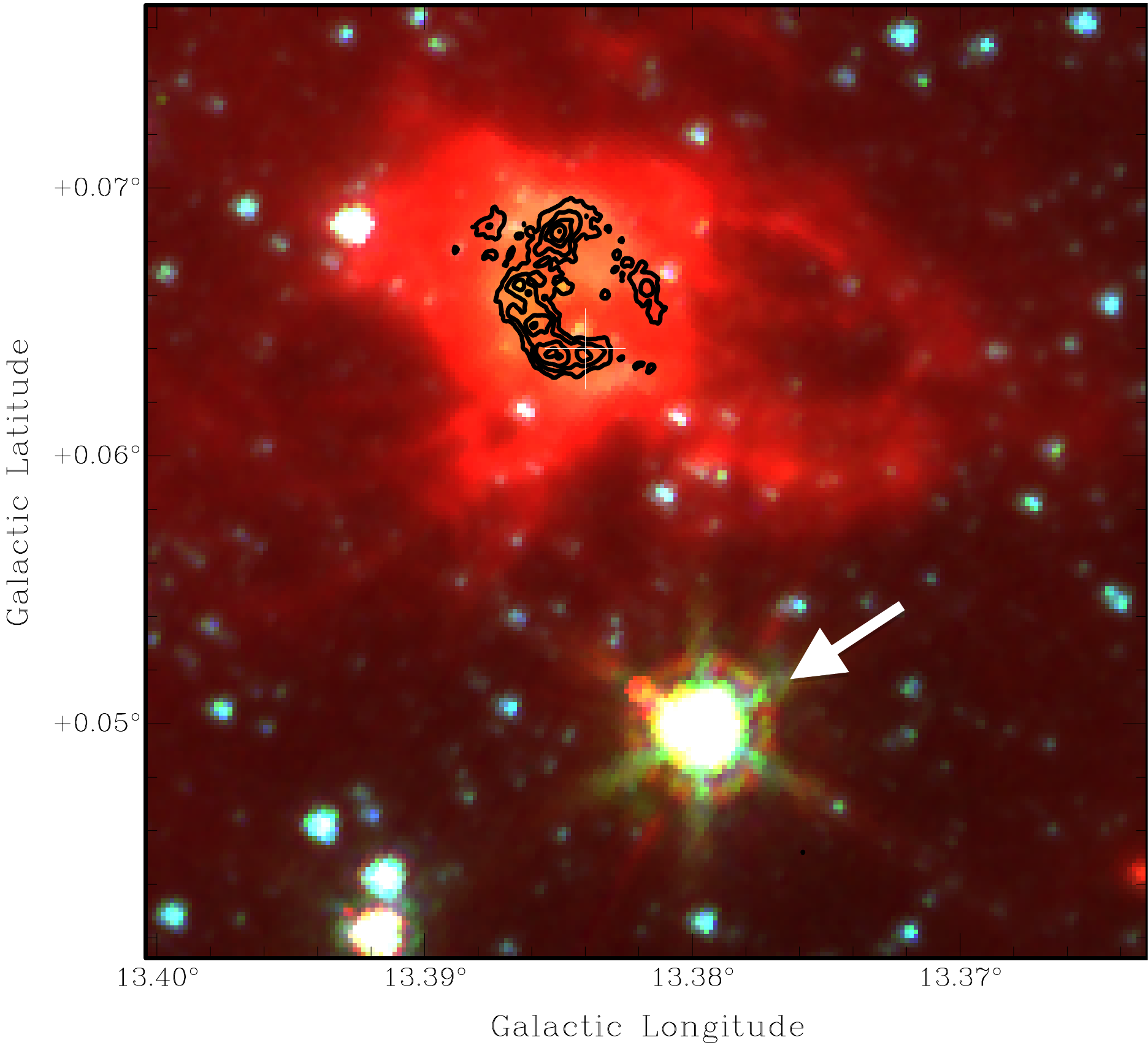}
	}
~\hfill
	\centering
	{%
	\includegraphics[width=0.48\textwidth]{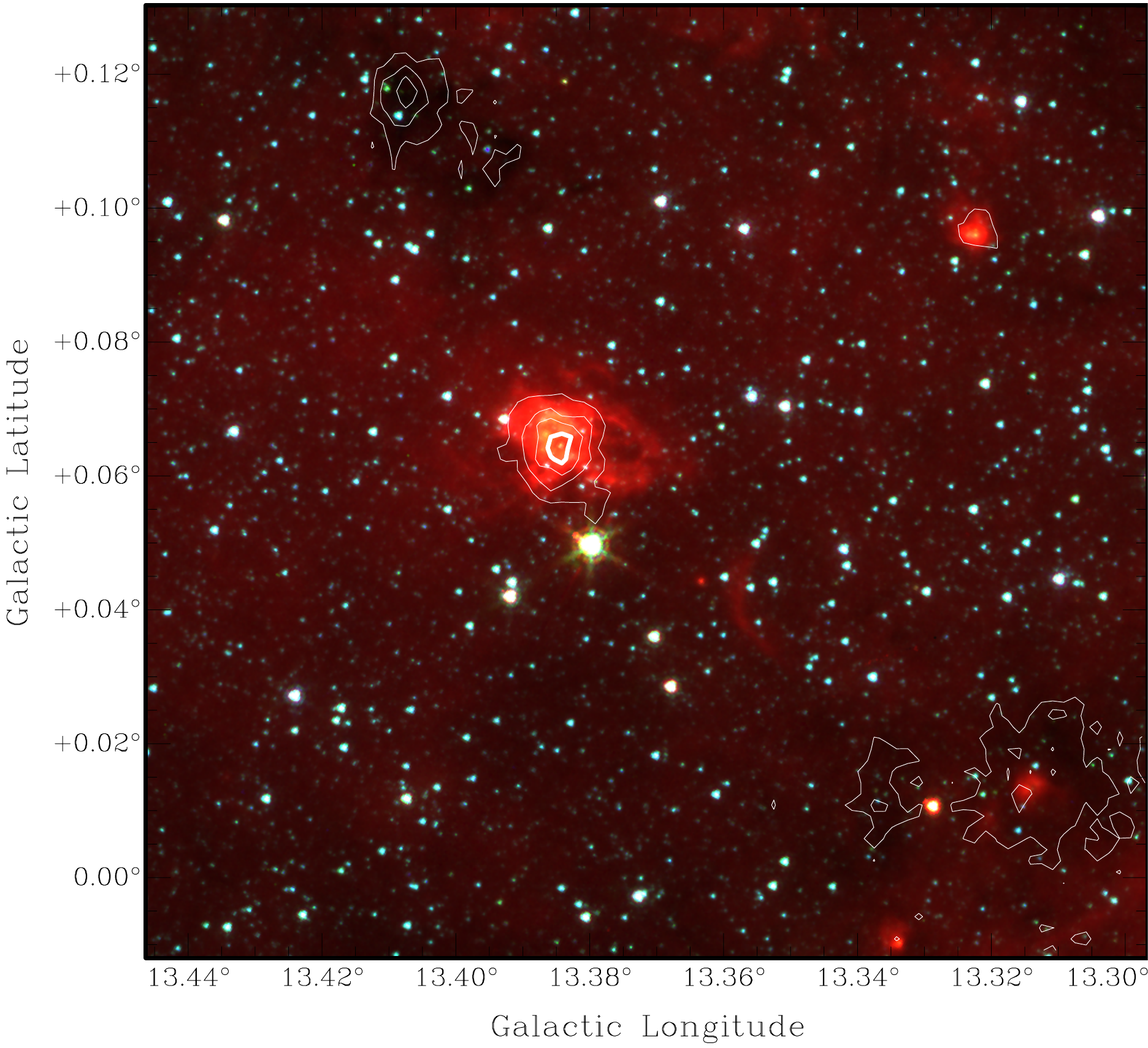}
	}
	\caption{GLIMPSE 3 colour image with blue as 3.6\,$\mu m$, green as 4.5\,$\mu m$ and and red as 8.0\,$\mu m$.  Contour image on the left shows CORNISH 5\,GHz radio continuum emission from 2\,mJy/beam to 10\,mJy/beam in steps of 2\,mJy/beam.  The white cross on the image on the left represents the centre of the ATLASGAL contours and the image on the right shows the GLIMPSE 3 colour image with ATLASGAL contours. The contour levels are from 0.25\,Jy/beam and increase in steps of 0.25\,Jy/beam up to 1.0\,Jy/beam.  We identify an evolved star, seen at G13.380+0.050 and discussed in section 3.8 that is unrelated to the source.}
	\label{GLIMPSE}
\end{figure*}

\section{Results}
\subsection{Infrared \& Radio Observation}
\label{sub:IR_radio}

The GLIMPSE \cite{Churchwell, Benjamin} and MIPSGAL \cite{Carey09} images of the star forming region G13.384+0.064 show a bright core surrounded by gas with an empty area or bubble between the two (Figure \ref{GLIMPSE}).  The bubbles are similar to those seen in other high mass star forming regions and they are likely to be created by stellar feedback (e.g. Weaver et al. 1977).  The CORNISH radio continuum image has sufficient resolution to show an extended shell like source in the centre of the region (Figure \ref{GLIMPSE}). The region inside the radio emission also contains low levels of diffuse emission in the UKIRT Infrared Deep Sky Survey (UKIDSS) K band images, as shown in Figure \ref{3cimage}. It is likely this emission corresponds to Brackett Gamma (B$\gamma$) emission associated with the H{\scriptsize II} region, as the morphology of the diffuse K band emission resembles the radio continuum contours \cite{Beck10}. 

We used data from UKIDSS \cite{Lawrence07}, 2MASS \cite{Skrutskie} and GLIMPSE \cite{Churchwell, Benjamin} to investigate the near and mid-IR source populations surrounding G13.384+0.064 and search for an embedded young stellar population.   First, we downloaded all the sources in the UKIDSS point source catalogue within a 10\,arc minute radius of G13.384+0.064 and looked for an increase in the surface density of infrared objects towards the source. There is no statistical difference in the surface density of infrared sources at the location of G13.384+0.064 compared to similar regions close to, but offset from the ATLASGAL contours. We conclude there is no evidence of a cluster based on source density.

\begin{figure}
	\includegraphics[width=0.5\textwidth]{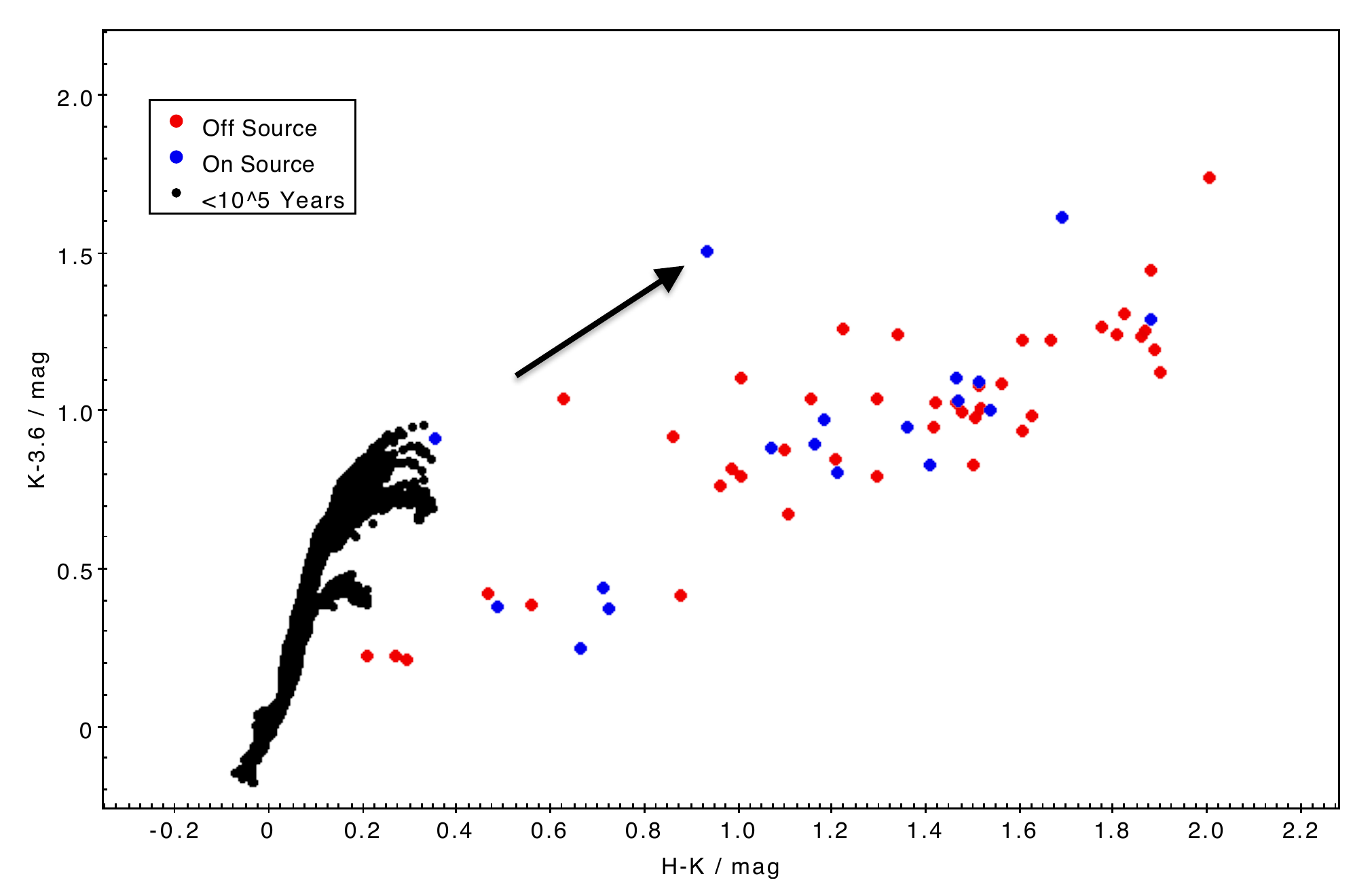}
	\caption{Colour-Colour diagram plotting sources 30 arc second around G13.384+0.064 (on source;blue) and 30 arc seconds around G13.36+0.075 (off source;red).  The IMF curve is data representing a cluster $<$10$^{5}$ years old from Lejeune \& Schaerer \protect\shortcite{Lejeune}.  The vector represents the extinction vector plotted as per Nishiyama et al. \protect\shortcite{Nishiyama}.  There are no clear signs there are a cluster of main sequence stars.}
	\label{cc}
\end{figure}

\begin{figure}
	\includegraphics[width=0.48\textwidth]{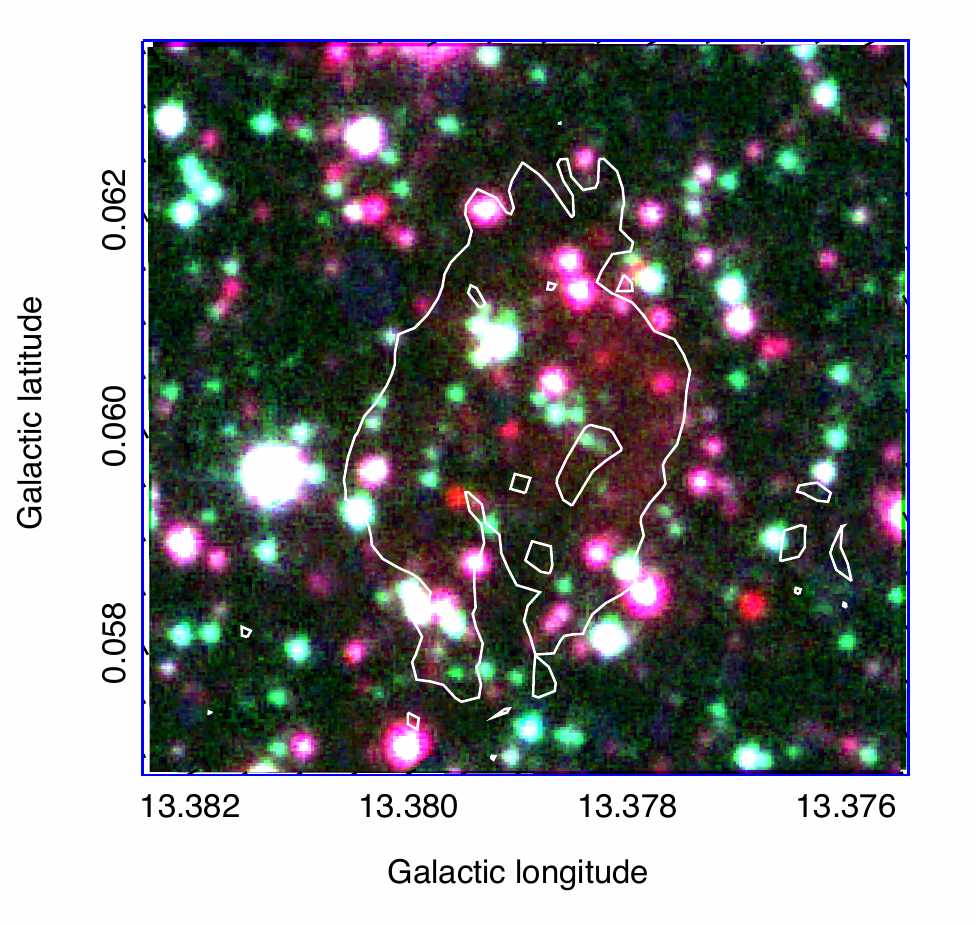}
	\caption{Three-colour image with red at K band, green at H band and blue at J band of UKIDSS data overlaid with contours from CORNISH 5\,GHz radio continuum emission with contours at 0.19 and 0.32\,Jy/beam.}
	\label{3cimage}
\end{figure}

The 3-colour diagram of UKIDSS data, Figure \ref{3cimage}, shows some red stars within the contours from CORNISH. To determine if these are field stars behind the dust cloud or if they are within the cloud, we use the UKIDSS, 2MASS (for saturated UKIDSS sources) and GLIMPSE photometry data to plot a colour-colour diagram (CCD) and colour-magnitude diagrams (CMDs) of the infrared sources on and off G13.384+0.064. Figure \ref{cc} shows a representative colour-colour diagram using H and K-band data from UKIDDS \cite{Lawrence07} and 2MASS \cite{Skrutskie} and L (3.6\,$\mu$m) band data from GLIMPSE \cite{Churchwell, Benjamin} for a 30 arc second radius around the source at G13.384+0.064 (matching with the contours from ATLASGAL shown in Figure 2) and off the source at G13.36$+$0.075, as shown in Figure \ref{cc}.  The reddening vector was determined from Nishiyama et al. \shortcite{Nishiyama} and the IMF data was plotted from Lejeune \& Schaerer \shortcite{Lejeune} for a cluster less than 10$^{5}$ years old.  While there are clearly reddened sources in the field, the CCD does not show an obvious excess of intrinsically red stars towards G13.384+0.064. The reddening value for each star was measured and a KS test was completed to see if the two populations of data (on and off source) were similar.  The results suggest that two populations could not be differentiated, making the red stars (from K band UKIDSS data) in Figure \ref{3cimage} likely to be field stars and not representative of a cluster. The only possible exceptions are the two blue dots at [K] $-$ [3.6] $\sim$1.5, which are potential candidates for YSOs of interest. However, without higher resolution data it is not possible to determine if these are simply highly extincted background stars or intrinsically red YSOs. Regardless of the nature of these two sources, the IR information from these 3 surveys data show no evidence of a sizeable distributed population of YSOs in the region.

We also note that these data may not be sensitive enough to detect stars deeply embedded into the natal cloud.  As described in Feldt et al. \shortcite{Feldt}, it might be possible to hide a cluster of smaller stars.  For further discussion see \S4.2.

\subsection{Distance}

The near kinematic distance of 1.9\,kpc and far distance of 14.1\,kpc  was measured by Schlingman et al. \shortcite{Schlingman} using HCO$^{+}$(3-2) and N$_{2}$H$^{+}$(3-2) observed by the Heinrich Hertz Submillimeter Telescope  at 1.1\,mm.  The velocity observed for these two molecules are 15.1\,km\,s$^{-1}$ for HCO$^{+}$ and 14.1\,km\,s$^{-1}$  for N$_{2}$H$^{+}$.  Lockman \shortcite{Lockman} reported 3\,cm radio recombination line observations completed using the Greenbank telescope with an observed $V_{LSR}$ of 18.3$\pm$1.8\,km\,s$^{-1}$.  However, Lockman \shortcite{Lockman} commented that the FWHM line for this source was so broad that the results could not be easily interpreted.  

H$_{2}$CO was observed at 4.83\,GHz, using the Nanshan station \cite{Du}. The results for the source G13.384 +0.064 show two molecular clouds.  The first cloud has a velocity of 10.71$\pm$0.40\,km\,s$^{-1}$ and the second has a velocity of 51.34$\pm$0.28\,km\,s$^{-1}$.  The velocity of the first cloud agrees well with the HCO$^{+}$,  N$_{2}$H$^{+}$ and radio recombination observations.  Data from the James Clark Maxwell Telescope (JCMT) reported three velocity components for $^{12}$CO(J=3-2) at 10, 15 and 50\,km\,s$^{-1}$  \cite{Dempsey13}.  It is likely, then, that the star formation is associated with the gas at 15\,km\,s$^{-1}$, rather than the 50\,km\,s$^{-1}$, since the gas at 50\,km\,s$^{-1}$ only appears in the relatively low density gas tracers of H$_{2}$CO and $^{12}$CO and the star formation is most likely associated with the denser gas.

The distance to this region has been determined by Schlingman et al. \shortcite{Schlingman} from the radial velocity and the Brand \& Blitz \shortcite{brand1993} Galactic rotation curve. An issue that affects all sources located inside the solar circle is that there are two kinematic solutions spaced at equal distances from the tangent position; these are commonly referred to as the \emph{near} and \emph{far} distances. These distance ambiguities can be resolved using H{\scriptsize I} data by comparing the velocity of absorption dips seen in the spectra with the source velocity as measured from thermally excited molecular lines. The two most commonly used methods are H{\scriptsize I} self-absorption (HISA; e.g., \cite{jackson2002, roman2009}) and H{\scriptsize I} emission-absorption (HIEA; e.g., Kolpak et al. \shortcite{kolpak03}, Andersen \& Bania \shortcite{andersen2009}, Urquhart et al. \shortcite{Urquhart12}). 

\begin{figure}
	\centering
	 {%
	 \includegraphics[width=0.5\textwidth]{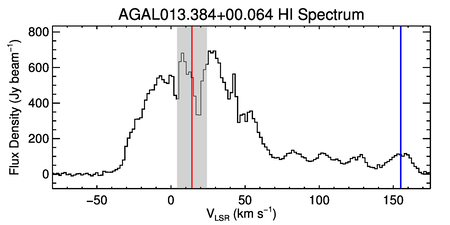}
	}
~\hfill
	\centering
	{%
	\includegraphics[width=0.5\textwidth]{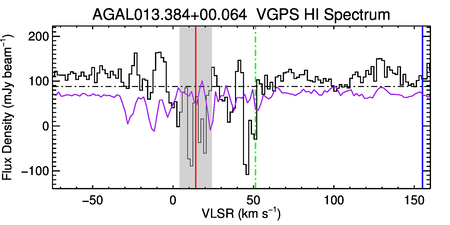}
	}
	\caption{The H{\scriptsize I} continuum subtracted (top) and H{\scriptsize I} continuum (bottom) profiles seen towards G13.384+0.064 and its associated H{\scriptsize II}.  In both of these panels, the source and tangent velocities are indicated by the red and blue vertical lines and the grey region shows a region 10\,km\,s$^{-1}$ either side of the source velocity. In the lower panel the green line indicates the maximum velocity found of the absorption features and the magenta line shows the 5$\sigma$ rms noise for the H{\scriptsize I}  data (see Urquhart et al. \protect\shortcite{Urquhart12} for more details). The presence of an absorption feature at a similar velocity as the source in the upper panel and the lack of absorption features up to the tangent velocity in the lower panel both strongly support and near kinematic distance for this source. }
	\label{HIabs}
\end{figure}

In Figure \ref{HIabs} we present the HISA profile (continuum subtracted) and HIEA (continuum included) H{\scriptsize I} profiles seen towards G13.384+0.064 and its associated H{\scriptsize II} region. The source velocity is approximately coincident with the velocity of a broad absorption feature seen in the H{\scriptsize I} profile; this is consistent with the source being located at the near distance due to the fact that there is too much intervening warm H{\scriptsize I} gas at the same velocity as the source for any absorption to be present for sources located at the far distance. 

The HIEA method is based on the principle that for any strong emission source located at the far distance we would expect to observe absorption features at all velocities up to and including the tangent velocity. This is due to the high density of cold H{\scriptsize I} clouds along any line of sight through the inner Galaxy. The lack of any absorption features between 50 and 150\,km\,s$^{-1}$ would suggest that the source is again located at the near distance. The absorption seen at 50\,km\,s$^{-1}$ may suggest that the source velocity may be incorrectly assigned to this source and may in fact be associated with another object within the line of sight. However, the lack of any emission from high-density molecular tracers (HCO$^{+}$ and N$_2$H$^{+}$) would rule out this possibility. Both methods therefore suggest a near distance is more likely.  For further discussion see \S \ref{far}.

The uncertainties in the near kinematic distance, allowing for $\pm$10\,km\,s$^{-1}$ when the streaming motions and peculiar velocities are considered, is $\pm$0.8\,kpc.  

\subsection{Dust Derived Mass}
The near distance of 1.9$\pm$0.8\,kpc was used by Miettinen \shortcite{Miettinen} and Urquhart et al.  \shortcite{Urquhart} to calculate the mass of the cloud surrounding the star as being 65$\pm$47 and 105$\pm$73\,M$_{\odot}$, respectively.  Urquhart et al. \shortcite{Urquhart} used the integrated flux measured in ATLASGAL \cite{Contreras2013} at 870\,$\mu$m and a temperature of 20\,K. Miettinen \shortcite{Miettinen} used 870\,$\mu$m observations from LABOCA and assumed a temperature of 35\,K. Since both reported masses are from the same instrument the differences are from the different temperatures and calibration errors.  By using the Hildebrand (1983) equation and making the same assumption as stated in equation 1 of Urquhart et al. \shortcite{Urquhart}, we determine the cloud mass to a greater accuracy from the derived values specific to this source.

\begin{equation}
\frac{M_{clump}}{M_{\odot}} = (\frac{D}{kpc})^{2} (\frac{S_{\nu}}{mJy}) \frac{R}{B_{\nu}(T_{dust}) \kappa_{\nu}}
\end{equation}

where $S_{\nu}$ is the integrated flux at 870\,$\mu$m of 603.94\,mJy \cite{Schuller09}, D is the heliocentric distance to the source (1.9$\pm$0.8\,kpc; \S 3.2), R is the dust-to-gas mass ratio (assumed to be 100), $B_{\nu}$ is the Planck function for a dust temperature $T_{dust}$(calculated in \S3.4 to be $33.7\pm1.5$\,K) and $\kappa_{\nu}$ is the dust absorption coefficient taken as 1.85\,cm$^{2}$\,g$^{-1}$ (as used by Urquhart et al. 2013).  This yields a value for the clump mass around the star as 57$\pm$35\,M$_{\odot}$. 

\subsection{Luminosity}
An estimation of the luminosity was determined from the IRAS fluxes \cite{Neugebauer} using the equation initially published by Casoli et al. \shortcite{Casoli} but using the same assumptions as equation 3 of Walsh et al. \shortcite{Walsh97}.  \\
\begin{equation}
F_{tot} = (f_{12}\delta\nu_{12}+f_{25}\delta\nu_{25}+f_{60}\delta\nu_{60}+f_{100}\delta\nu_{100})/0.61\\
\end{equation}
This yielded a value of $7.2\times10^{3}$\,L$_{\odot}$. This value is similar to that reported by Miettinen \shortcite{Miettinen} of $7.4\times10^{3}$\,L$_{\odot}$,  which also uses the IRAS flux, but using the same assumptions as Casoli et al. \shortcite{Casoli}. The bolometric luminosity reported by Urquhart et al. \shortcite{Urquhart}, calculated by scaling the MSX 21\,${\mu}m$ flux, was $3.4\times10^{3}$\,L$_{\odot}$  which is almost half the value calculated from the IRAS fluxes.  We note that the luminosity values reported by Miettinen \shortcite{Miettinen}, Urquhart et al. \shortcite{Urquhart} and contained in this work, all assume a distance of 1.9\,kpc.

The IRAS measurements can be considered as a strong upper limit on the bolometric luminosity. The large IRAS beam encapsulates all emission and there is no loss of flux due to extended emission that is masked out with other observations employing nodding or jittering method, such as BOLOCAM or ATLASGAL.

\begin{figure}
	\centering
	 {%
	 \includegraphics[width=0.51\textwidth]{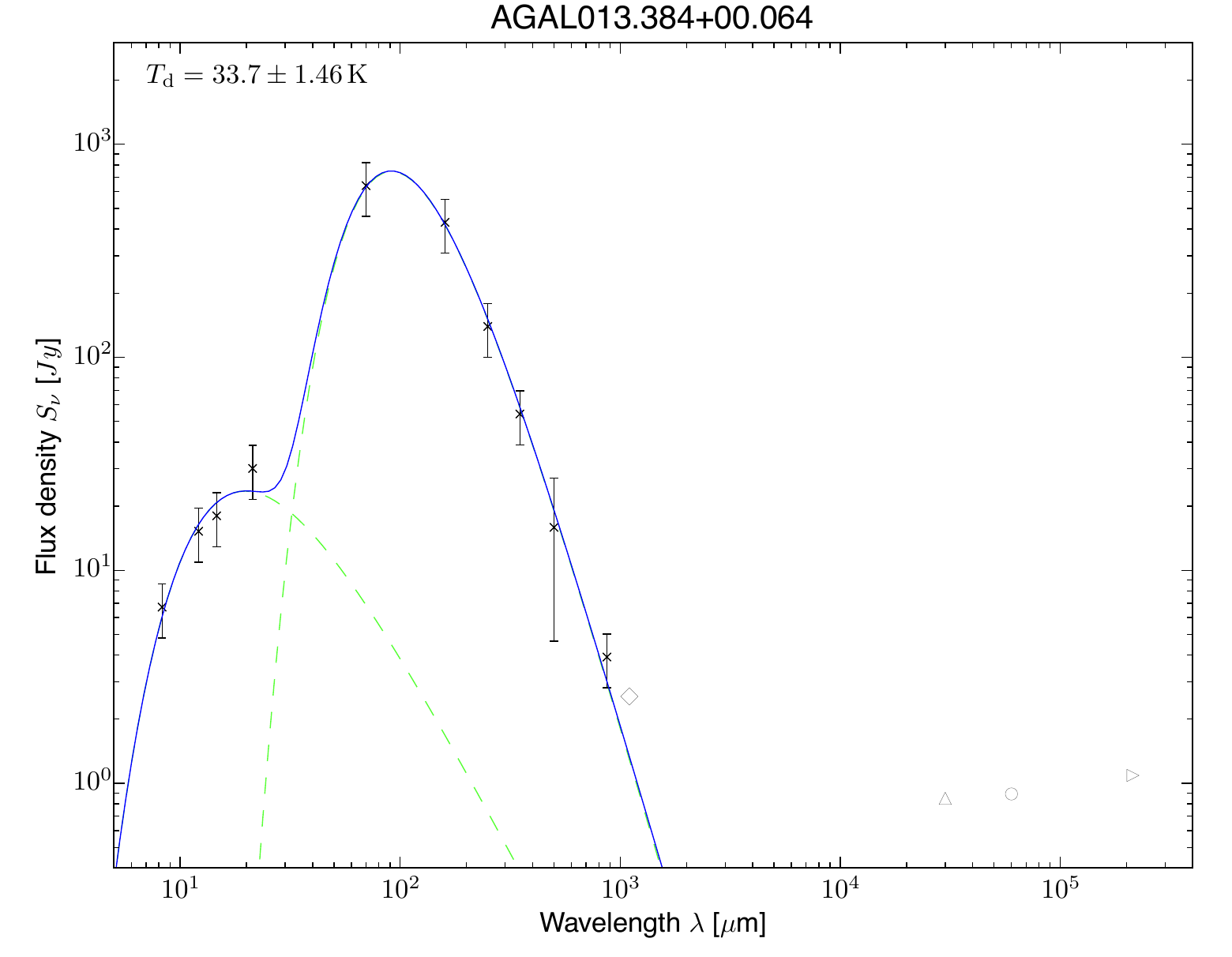}
	}
	\caption{Spectral energy distribution created by data of different surveys plotted on a log scale.  To obtain the dust temperature and luminosity, a two-component model was fitted to the flux densities  measured through aperture photometry from the MSX \protect\cite{Price2001}, Hi-GAL \protect\cite{Molinari2010} and ATLASGAL \protect\cite{Schuller09} maps (blue fit).  Additionally, catalogued data of the extended emission from the radio as well as some select infrared and submilimetre data.  
	Open Triangle-MSX \protect\cite{Egan03},
	Diamond-BOLOCAM \protect\cite{Schlingman},
	Triangle-Nobeyama \protect\cite{Handa},
	Open Circle-VLA 5\,GHz \protect\cite{Becker}, and
	Triangle Right-VLA 1.5\,GHz \protect\cite{Garwood}.}
	\label{light}
\end{figure}

A more accurate value of the bolometric luminosity is measured using two component fitting of the spectral energy distribution (SED).  The integrated flux data was compiled from a combination of reported catalogues values ranging from 4\,$\mu$m to 21\,cm as well as measured values through aperture photometry from the MSX \cite{Price2001}, Hi-GAL \cite{Molinari2010} and ATLASGAL \cite{Schuller09} maps.  A plot of the integrated fluxes versus wavelength shows a curve that peaks in the IR and is flat at radio wavelengths as shown in Figure \ref{light}.  This curve is typical of the spectral energy distribution of an embedded forming high mass star.  The peak in the infrared emission is characteristic of the short wavelength stellar light being reprocessed to longer wavelengths by dust in the surrounding gas cloud, and the flat region is from free-free radio emission characteristic of an H{\scriptsize II} region. 

The flux density from the MSX, Hi-GAL and ATLASGAL maps, Fapp, was
measured within an aperture radius of 35.7Ó, corresponding to 3$\sigma$
of a Gaussian fitted to the source and centred on the peak flux pixel
position of the 250\,$\mu$m image. The background flux density, Fbg,  was
obtained as the median pixel value from an annulus with r$_{inner}$=47.6
arcsec to r$_{outer}$=59.5 arcsec around the aperture. Subtracting the
background flux density from the aperture flux density yields the
background corrected source flux $F = Fapp - Fbg$, reconstructing the
SED in 10 bands from 8\,$\mu$m to 870$\mu$m. The errors of the flux densities
are calculated by adding the absolute calibration uncertainty to the
intrinsic measurement error in quadrature. We assume a measurement
uncertainty of 20\,per\,cent for all bands except for the 500\,$\mu$m band,
where we take into account the large pixel size of 15\,arcseconds,
hence assuming a measurement uncertainty of 50\,per\,cent.

The SED was then fitted with a two-component model consisting of a greybody and blackbody. The greybody (i.e. a modified blackbody) models the cold dust envelope's emission, taking into account the wavelength dependence of the dust in the far-infrared to submm wavelength regime, whereas the blackbody models a hot, optically thick, deeply embedded component:  
 \begin{multline}
F_\lambda(T_\mathrm{d}, \beta, \tau_{870}, T_\mathrm{h}, \Omega_\mathrm{h}) = \\ F_{\lambda, \mathrm{hot}}(T_\mathrm{h}, \Omega_\mathrm{h}) + F_{\lambda, \mathrm{dust}}(T_\mathrm{d}, \tau_{870})
\end{multline}
where $F_{\lambda, \mathrm{hot}}$ is the hot component given by a blackbody scaled with the effective solid angle of the hot component and $F_{\lambda, \mathrm{dust}}$ is the greybody emission from the dust envelope given by:
\begin{multline}
F_{\lambda, \mathrm{dust}}(T_\mathrm{d}, \tau_\mathrm{870})= \\ \Omega_\mathrm{d}\cdot B_\lambda(T_\mathrm{d})\cdot \left(1-e^{-\tau_\mathrm{870} \left(\frac{870\mathrm{\mu m}}{\lambda}\right)^\beta}\right)
\end{multline}
where $\Omega_\mathrm{d}$ is the solid angle subtended by the source, $B_\lambda(T_\mathrm{d})$ the blackbody intensity at the dust temperature $T_\mathrm{d}$, $\tau_\mathrm{870}$ the dust optical depth at the reference wavelength of 870\,$\mu$m and $\beta$ the dust spectral index. We leave the dust spectral index $\beta$ fixed to a value of 1.75, as computed as the mean value from the dust opacities of Ossenkopf \& Henning \shortcite{Ossenkopf1994} for the submm regime. As a result the dust temperature was determined to be $33.7\pm1.5$\,K, which is consistent with the result used by Miettinen \shortcite{Miettinen} and a luminosity of 4.1($\pm$1.7)$\times$10$^{3}$\,L$_{\odot}$ which is comparable to that reported by Urquhart et al. \shortcite{Urquhart} and is consistent with the upper limit from the IRAS measurements .

\subsection{Lyman Continuum Flux}
A number of observations of the region were completed in the radio using the NRAO Very Large Array (VLA) \cite{Zoon, Becker, Garwood, Purcell13}, Nanshan Radio Telescope \cite{Du}, NRAO Green Bank (GBT), \cite{Lockman} and the Effelsberg 100'm Telescope \cite{Altenhoff}. The integrated flux from two 6\,cm observations, CORNISH \cite{Purcell13} and VLA 5\,GHz Survey \cite{Becker}, were used to calculate the Lyman-continuum flux based on equation 1 and 3 in Kurtz et al.  \shortcite{Kurtz94}, which represent modified equations presented by Mezger \& Henderson (1967).  Both of these surveys were completed using the NRAO VLA but CORNISH used B configuration,with a restoring beam of 1.5$^{\prime\prime}$, which is not as sensitive to the extended emission as the survey done by Becker et al. \shortcite{Becker} which was conducted in C configuration and has a restoring beam of 4$^{\prime\prime}$. Therefore, we expect the C-array data to recover more emission. The logarithm of the number of Lyman continuum photons per second (Log N$_{c}$) using the CORNISH integrated flux of 603\,mJy and an assumed gas to dust ratio of 100, was determined to be 47.3\,photon s$^{-1}$,  as reported by Urquhart et al. \shortcite{Urquhart}. Using the integrated flux from Becker et al. \shortcite{Becker} of 891.8$\pm$20.9\,mJy, Log (N$_{c}$) = 47.3$^{+0.2}_{-0.4}$\,photon\,s$^{-1}$.  

The flux derived by Altenhoff et al. \shortcite{Altenhoff} for the 100\,m Effelsberg telescope at 4.9\,GHz is 0.9\,Jy, which is similar to the value reported by Becker et al. \shortcite{Becker} of 0.891\,Jy.  This suggests that all the extended emission was accounted for in the C configuration observations.

\subsection{Star Type}
The Lyman-continuum flux and bolometric luminosity are compared to Table 1 in Davies et al. \shortcite{Davies11} to determine a mass of the star powering the H{\scriptsize II} region.  The bolometric luminosity as determined by the SED fit is Log(L$_{\star}$/L$_{\odot}$) = 3.61$\pm$0.14.  This relates to a star between 9-12\,M$_{\odot}$.  The Log of the Lyman photon flux  is 47.3$^{+0.2}_{-0.4}$\,photon\,s$^{-1}$.  This relates to a star between 15-20\,M$_{\odot}$.  

The mass measured using the SED fit is 13$\pm$6\,M$_{\odot}$.  A main sequence star type B, is a star with a mass of 2.1-16\,M$_{\odot}$ which is consistent with the values measured for this source. As B stars are known to have an excess of Lyman flux (see $\S$ 4.1), the star is likely to be between 9-12\,M$_{\odot}$ as suggested by the bolometric luminosity measurement (as suggested by Table 1 in Davies et al. \shortcite{Davies11} and SED curve in Figure 6).

\subsection{An unrelated evolved star}
Figure \ref{GLIMPSE} shows a bright infrared star at G13.380+0.050, approximately one arcminute to the south. This star coincides with a 1612\,MHz OH maser, detected by Sevenster et al. \shortcite{Sevenster}, indicating that the star is evolved and unlikely to be part of the same star formation process. Given the offset between this star and G13.384+0.064, the projected distance is 0.5\,pc. The image in Figure \ref{GLIMPSE} shows the structure of the cloud in the infrared extends toward the star at G13.380+0.050 but does not overlap with this star. Furthermore, there is no edge brightening in the extended infrared emission observed. This suggests that the bright star is unlikely to have a direct physical influence on the star forming region and is most likely an unrelated star projected along the line of sight, but not necessarily physically close to the star forming region.

\section{Discussion}

\subsection{Far Distance Consideration}
\label{far}

The radio continuum-derived luminosity of  $3.6~\times~10^{4}\,{\rm L}_\odot$ derived in $\S$3.5 is a factor of ten greater than the dust-derived luminosity $4.1\times10^{3}\,{\rm L}_\odot$. We must reconcile this difference.

If we assume that G13.384+0.064 is at the far kinematic distance of 14.1\,kpc, then the radio-derived and dust-derived luminosities are $3.2\times10^5\,{\rm L}_\odot$ and $2.2\times10^{5}\,{\rm L}_\odot$, respectively. These two luminosities are more consistent than those when the near distance is assumed. However, even though this luminosity argument favours the far kinematic distance, we still favour the near kinematic distance for the following reasons. 

Firstly, as discussed in $\S$3.2, the H{\scriptsize I} observations strongly favour the near distance. The H{\scriptsize I} observations analysed by Urquhart et al. \shortcite{Urquhart} and shown in Figure \ref{HIabs} shows little evidence of high velocity H{\scriptsize I} self-absorption, which would be expected from intervening gas, if it was at the far kinematic distance. The only high velocity absorption is seen around 50\,km\,s$^{-1}$. This corresponds to the velocity of the cloud seen in the H$_2$CO and $^{12}$CO observations reported previously and so the absorption is most likely not related to G13.384+0.064.

Secondly, there are uncertainties in the expected Lyman continuum flux from young high mass stars. Smith \shortcite{Smith13} concludes that the Lyman continuum flux of early B type stars may be in excess of previous models, caused by the accretion of cold gas from the circumstellar disk onto hot-spots on the surface of the young star. Comparing the measurements for G13.384+0.064, we find that it sits on the upper edge of their model such that the excess radio continuum flux is consistent with the bolometric luminosity. If the radio continuum flux were much larger, this source would show an unaccounted for excess of radio continuum, compared to the dust-derived luminosity.

A final consideration is the expected number of lower mass stars surrounding the high-mass star if it were at the near vs far kinematic distance. If G13.384+0.064 was at the far distance of 14.1\,kpc, then the mass of the star would be approximately 38\,M$_{\odot}$, based on the mass to luminosity relationship. The mass of the surrounding gas cloud would increase to 5800\,M$_{\odot}$, making it very likely that a cluster is present. However, as discussed in $\S$~\ref{sub:IR_radio}, no evidence for a cluster is seen in the infrared images.

\subsection{Isolated vs. clustered star formation in G13.384+0.064}
\label{sub:isol_vs_cluster}

We now seek to use the properties derived above to determine whether G13.384+0.064 
is indeed a high-mass star forming in isolation. 

Parker et al. (2007) defines
an isolated B type star as a star 10\,M$_{\odot} < {\rm M}_* < 17.5 {\rm M}_{\odot}$ in which the cluster mass is $<100\,{\rm M}_{\odot}$ and there are no O-stars present. The estimated mass of the star based on the SED
curve is 13$\pm$6\,M$_{\odot}$ and the estimated remaining clump mass is
57$\pm$35\,M$_{\odot}$. Both of these are consistent with
the definition described by Parker et al. (2007) for isolated
high mass star formation.

Based solely on the detection of a strong radio continuum source, we can conclude that there must be at least
one embedded high mass star. Incorporating a model by Smith \shortcite{Smith13}, we find that such a star must account for
nearly all (if not, all) of the infrared flux, leaving little room
for a cluster. This is because lower mass stars will not contribute significantly to the radio-derived luminosity, but will
significantly contribute to the infrared or dust-derived luminosity.

We can model what a cluster might look like, by constraining the
quantities of dust-derived luminosity and Lyman continuum
flux. In order to do this, we use a Monte-Carlo simulation
of the initial mass function (IMF) to generate a cluster.
Based on previous work by Walsh et al. (2001), we found that
commonly-used functions of the IMF do not greatly affect
the make-up of simulated clusters, so we choose the IMF
model of Kroupa et al. (1993). In our simulation, we randomly generate stars between masses of 0.1 and 100\,M$_\odot$,
according to this IMF and then calculate the cluster physical parameters, such as total mass, luminosity and Lyman
continuum flux. In order to measure the relationship between Lyman continuum flux and luminosity, we use the
values given in Table 1 of Thompson (1984) but we note
that previous studies have found an excess of Lyman continuum photons from early B-type stars (eg. Urquhart et al. 2013). Therefore, in order to take into account the work of Smith (2014) we apply their most extreme case for ratio of Lyman flux, compared to previous models, where the Lyman flux may be reduced by up to an order of magnitude for the same star with the same bolometric luminosity (ie. reduction of 10$^{47.3}$ to 10$^{46.3}$ photon s$^{-1}$). It is important to note that in taking such an extreme reduction in Lyman flux, our simulation will favour the formation of a cluster, rather than an isolated star. This allows us to use the data of Thompson (1984) and correct it for more recent modelling by Smith (2014).

In our Monte-Carlo simulation, we continue to add members to the cluster
until the total cluster luminosity is greater than $4.1 \times
10^3$\,L$_\odot$. We generated 68,877 clusters with sufficient luminosity
to meet this criterion. However, we note that the majority of generated
clusters have total luminosities far in excess of this value. This is
because the last star added to the cluster is typically a high mass star
with very high luminosity. Thus, we exclude those generated clusters that
have luminosities in excess of the IRAS-derived luminosity ($7.2 \times
10^3$\,L$_\odot$), leaving 30,601 clusters. We choose the IRAS-derived
luminosity here because it is a strong upper limit on the bolometric
luminosity, given that IRAS will likely overestimate the total infrared
flux, but not underestimate it.

Of our remaining clusters, we find that the median luminosity for the highest
mass star in each cluster is $2.8 \times 10^3$\,L$_\odot$ which means
that for most clusters, the luminosity is dominated by one star. We also
find that the highest mass star generated in any cluster has a luminosity
of $7.1 \times 10^3$\,L$_\odot$, which we calculate has a corresponding
Lyman continuum flux of $10^{45.7}$\,photons\,s$^{-1}$. This Lyman flux
is lower than we expect ($10^{46.3}$ photons\,s$^{-1}$) by about an order
of magnitude. In summary, our simulations indicate that it is very
difficult to randomly generate a cluster with the properties that we observe for G13.384+0.064.
The only way to generate a good match is for the first star
selected from the IMF to be a high mass star with the right luminosity and
Lyman flux properties. However, classifying such a
single star as a cluster is questionable.

\section{Conclusions}
In this paper we compared the Lyman-continuum photon flux and clump mass of approximately 200 star forming regions to search for high mass stars forming in isolation.  We identified one source, G13.384+0.064, as a very promising candidate. Analysis completed in both the infrared and the radio, combined with Monte Carlo modelling, shows this source is consistent with a single high mass star in formation, and there is no strong evidence of a cluster. This verifies the choice of using Lyman flux and bolometric luminosity to identify such candidates. While the observations in the literature allowed us to rule out a large population of embedded young stars, further high-resolution, deep, infrared and sub-mm observations are required to quantify just how small this population is, and thereby unambiguously determine if G13.384+0.064 is a high mass star forming in isolation.

\section*{Acknowledgements}
The authors would like to thank the referee for the comments and advice. SNL would like to thank Andy Longmore for very helpful discussions about the analysis of the infrared data. This research has made use of the SIMBAD database, operated at CDS, Strasbourg, France. This research has also made use of the VizieR catalogue access tool, CDS, Strasbourg, France. The original description of the VizieR service was published in A\&AS 143, 23.

\label{lastpage}

\end{document}